\documentclass[preprint]{elsarticle}

\usepackage{graphicx}%
\usepackage{multirow}%
\usepackage{amsmath,amssymb,amsfonts}%
\usepackage{amsthm}%
\usepackage{mathrsfs}%
\usepackage[title]{appendix}%
\usepackage{xcolor}%
\usepackage{textcomp}%
\usepackage{manyfoot}%
\usepackage{booktabs}%
\usepackage{algorithm}%
\usepackage{algorithmicx}%
\usepackage{algpseudocode}%
\usepackage{listings}%

\theoremstyle{thmstyleone}%
\newtheorem{theorem}{Theorem}%

\newcommand{\Cov}{\text{Cov}}
\newcommand{\Var}{\text{Var}}
\newcommand{\ESS}{\text{ESS}}
\newcommand{\IACT}{\text{IACT}}
\newcommand{\scaleval}{0.55}

\begin{document} 

\begin{frontmatter}

\title{Estimating the Effective Sample Size for an inverse problem in subsurface flows}

\author{Lucas Seiffert\corref{cor1}}
\ead{lucas.seiffert@utdallas.edu}
\cortext[cor1]{Corresponding author}
\author{Felipe Pereira}
\ead{luisfelipe.pereira@utdallas.edu}
\affiliation{organization={Department of Mathematical Sciences,The University of Texas at Dallas},
             addressline={800 W. Campbell Road},
             city={Richardson},
             postcode={75080},
             state={Texas},
             country={United States}}

\begin{abstract}
The Effective Sample Size (ESS) and Integrated Autocorrelation Time (IACT) are two popular criteria for comparing Markov Chain Monte Carlo (MCMC) algorithms and detecting their convergence. Our goal is to assess those two quantities in the context of an inverse problem in subsurface flows. We begin by presenting a review of some popular methods for their estimation, and then simulate their sample distributions on AR(1) sequences for which the exact values were known. We find that those ESS estimators may not be statistically consistent, because their variance grows linearly in the number of sample values of the MCMC. Next, we analyze the output of two distinct MCMC algorithms for the Bayesian approach to the simulation of an elliptic inverse problem. Here, the estimators cannot even agree about the order of magnitude of the ESS. Our conclusion is that the ESS has major limitations and should not be used on MCMC outputs of complex models. 
\end{abstract}


\end{frontmatter}

\section{Introduction}

In the study of subsurface flows, one may be interested in obtaining the properties of the porous medium given some measurements of the flow. This can be modelled as the inverse problem of a partial differential equation. Since there is typically not enough data to fully characterize the solution, this problem is considered ill-posed. One way to approach it is by Bayesian Analysis: we choose a prior distribution for the quantity of interest and then find its posterior distribution conditioned on the available data.

The resulting distribution is usually complicated, so its simulation requires Markov Chain Monte Carlo (MCMC) methods. We recall that MCMC is the composition of two ideas: a Markov Chain is used to simulate values of a sequence of random variables whose distribution converges to the desired distribution. Then, those values are used to perform a Monte Carlo integration, to find the expected value of this distribution.

Analogously, the output analysis of an MCMC procedure can also be divided into the convergence of the Markov Chain and then its Monte Carlo part. For example, a popular criterion for the convergence of the Markov Chain is the Potential Scale Reduction Factor (PSRF) \cite{gelman2013bayesian}. Once the Chain has converged, we assess the quality of the Monte Carlo approximation. Here, since Markov Chains produce correlated sample values by their very definition, the usual Central Limit Theorem has a modified version with a different asymptotic variance. The concept of Effective Sample Size (ESS) comes from comparing this new variance with the variance that would be obtained by using i.i.d. sample values. 

Estimating the ESS of the output serves two main purposes. First, to obtain a confidence interval for the posterior mean, when a Central Limit Theorem is available. Secondly, to compare different methods in terms of how correlated their output is. The idea is that less correlated outputs are more efficient, by requiring a smaller sample size to reach a desired precision. 

Examples of recent use of the ESS for comparisons of MCMC methods can be found in \cite{lykkegaard2020multilevel, girolami2011riemann,tong2020mala, cotter2013mcmc, dodwell2019multilevel}. However, there are many different approaches to estimate this quantity, and the justification of why one method was preferred over another is rarely given. Moreover, we find ourselves in the very case in which estimating the ESS is difficult: as it will be shown in Section~\ref{simulations}, our chains are highly autocorrelated.

As it will be described in Section~\ref{MCMCtheo}, the ESS is the number of iterations performed so far divided by an infinite sum of quantities called autocorrelations. However, it is known that the estimation of these autocorrelations is tricky, especially when a sample is too correlated \cite{priestley1981spectral}.  This makes it challenging to study the properties of estimators for the ESS, and the few theoretical results available on them require strong assumptions \cite{madras1988pivot, vats2019multivariate}. As this quantity is gaining popularity to justify convergence of an MCMC algorithm, a natural question to ask is how reliable it actually is.

Our goal here was to test the methods cited in recent papers to calculate the ESS, and compare their results. What we have found is that these results may not even agree on the order of magnitude. Even more, we have found that, in practice, the ESS is not even a consistent statistic. Thus, blindly applying one ESS estimator to the output of an MCMC method is not a rigorous procedure to assess its convergence.

We have focused on the estimate of the ESS for only one parameter, that is, for a univariate distribution. There are generalizations of the ESS for multivariate target distributions \cite{roy2020convergence,vats2019multivariate}, but as it will be shown here, the univariate case is still challenging enough and does not produce satisfactory results. 

Finally, the statistical community also discusses more sophisticated and effective methods to estimate how fast the chain is mixing and how it is converging, called regeneration methods \cite{jones2006fixed}. They are, however, only proven to be valid for a finite discrete state space MCMC so far, and are thus not useful for inverse PDE problems yet. 

This paper is organized as follows. In the first few sections, a description of the inverse problem at hand is given, with emphasis on the framework of Bayesian Analysis. Then, we review some concepts of MCMC to introduce and discuss the theory behind the Effective Sample Size, including its uses and traditional formula. Then, we move on to describing the methods we compared, and test them on AR(1) sequences of a known specified autocorrelation. Finally, we show results for the inverse problem in subsurface flows, and summarize our conclusions and suggestions at the end of the paper.

\section{An elliptic problem in subsurface flows}\label{elliptic} 

The problem we're considering is a steady-state horizontal flow on a square, $\Omega = [0,1]\times[0,1]$. Assumptions for when this is a good model can be found in \cite{bear1987modeling}. 

The rock matrix effects on the flow of a fluid passing through the medium are described by a \textit{permeability field} $\kappa\colon\Omega\rightarrow R$, whose only constraint is that it be non-negative. The flow itself is described by an \textit{average flow velocity} field $\mathbf{u}$, its pressure field $p$, and a viscosity $\mu$. To simplify the model, we assume that $p$ does not depend on time, so that the flow is steady. Then, by Darcy's Law, 
\[\mathbf{u}(x,y) = -\frac{\kappa(x,y)}{\mu}\nabla p(x,y), \qquad (x,y)\in\Omega.\]
Assuming also that $\mu$ is a constant, the principle of Conservation of Mass can be applied,
\[\nabla\cdot\mathbf{u}=q, \]
where $q$ is a source term. This will imply the elliptic equation
\[-\nabla\cdot(\kappa\nabla p) = q.\]
To simplify further, we assume that $q=0$. We apply Dirichlet boundary conditions $p=1$ on the left edge $\Omega_L$, $p=0$ on the right edge $\Omega_R$; and no-flow Neumann conditions on top and bottom $\Omega_{TB}$. The full equation becomes
\[\left\{\begin{array}{l l}
 \nabla\cdot(\kappa\nabla p) = 0&\text{in }\Omega \\
 p = 0 &\text{on }\Omega_L\\
 p = 1 &\text{on }\Omega_R\\
 (\kappa\nabla p) \cdot \mathbf{n} = 0 &\text{on }\Omega_{TB}
 \end{array}\right.\]

\begin{figure}[h!]
\centering
\includegraphics[scale=0.6]{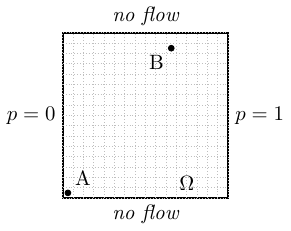}
\caption{Problem domain}
\end{figure}

A direct problem using this equation is to obtain the pressure $p$, when the parameter field $\kappa$ is known. Our interest, however, is in the inverse problem: given some measurements of the pressure, we want to estimate the permeability field $\kappa$. 

Realistic inverse models account for the the noisiness of the data. To keep the discussion simple, we assume that our problem has already been discretized, so that $\kappa$ is a vector of permeability values on points of a uniform grid, and $p$ is likewise a vector of pressure measurements. The Gaussian noise is incorporated as \cite{stuart2010inverse} 
\[p_{\text{obs}}=F(\kappa)+\epsilon,\qquad \epsilon\sim N(\mu, \Sigma), \]
where F is the solution of the forward problem with the indicated $\kappa$, and $\Sigma= I\sigma^2$ the \textit{precision matrix}. In our simulations, we fixed $\sigma^2= 1\times 10^{-3}$.  

\begin{figure}[h!]
\centering
\includegraphics[scale=0.6]{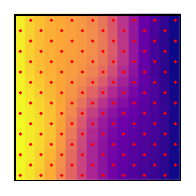}
\caption{Pressure data}
\label{fig:data}
\end{figure}

In this work, we use noisy data for the pressure in a chessboard pattern of the discretized $16\times 16$ grid (Figure \ref{fig:data}). This amount of data is not enough to determine a unique solution for the inverse problem, so it is ill-posed. 

This problem was first considered in \cite{morzfeld2015parameter}, but with pressure measurements on every cell of the discretized grid. The configuration from our work comes from \cite{tong2020mala}.

\subsection{The Bayesian approach to the inverse problem}\label{bayesian}

The idea of the Bayesian approach is to start with a probability distribution for the permeability field, and use Bayes' Theorem to incorporate the available data.  The resulting distribution is called the \textit{posterior distribution}, and it will indicate which values will be more likely to be the true ones, given all the information that we have.  We break down the explanation of this approach in steps, and refer to \cite{stuart2010inverse} for a more sophisticated discussion.  

\vspace{0.3cm}
\noindent\textbf{Step 1.} We must choose a prior distribution for the permeability field $\kappa$. The choice will be a lognormal field \cite{christakos1992random}, i.e., $\log (\kappa) = \eta$, where $\eta$ is a Gaussian field. It is determined by a mean field, chosen to be zero in our work, and a covariance kernel
\[K(x_1, x_2, y_1, y_2) = \exp\left(-\frac{(x_1-x_2)^2}{l_x^2}-\frac{(y_1-y_2)^2}{l_y^2}\right),\]
which models how the field is spatially correlated. It is parametrized by the \textit{correlation length scales} $l_x$ and $l_y$, and chosen to be $l_x=l_y=
0.2$ in our simulations.

\vspace{0.3cm}
\noindent\textbf{Step 2.} Let $p_{\text{data}}$ denote the pressure data. By \textit{Step 1}, we have a prior distribution $\pi(\eta)$. The way to incorporate $p_{\text{data}}$ to this distribution is by Bayes Rule, 
\[\pi(\eta|p_{\text{data}}) \propto l(p_{\text{data}}|\eta)\pi(\eta),\]
where $l$ is the likelihood function that arises from considering the distribution of the noise in the data as in the previous section,
\[l(p_{\text{data}}|\eta) \propto \exp\left\{-(p_{\text{data}}-F(\eta))^T\Sigma^{-1}(p_{\text{data}}-F(\eta)\right\}.\]

\vspace{0.3cm}
\noindent\textbf{Step 3.} This is an infinite-dimensional model, and it becomes a finite dimensional one when we discretize the domain $\Omega$. Still, the discretization will give rise to some $256$ parameters, which is a large simulation. The Karhunen-Lo\`eve decomposition will be used to reduce this field into a linear combination of the 20 most relevant eigenvectors of the Gaussian Field \cite{ali2021multiscale,ghanem2003stochastic}.

\vspace{0.3cm}
\noindent\textbf{Step 4.} The final task is to estimate $E\{\pi(\eta|p_{\text{data}})\}$. Since this is a complicated distribution, the estimation is done by Markov Chain Monte Carlo. 

\section{The Effective Sample Size}\label{MCMCtheo}

Let $\{X_n\}$ be a Markov Chain with stationary distribution $\pi(\eta|p_{\text{data}})$. Using one of the methods from \textit{Section}~\ref{mcmcmethods}, we can construct a realization of $N$ values of such a chain, denoted here $\{x_1, x_2, \ldots, x_N\}$. Their sample mean is given by
\[\overline{X}_N = \frac{1}{N}\sum_{t=1}^Nx_t.\]

By the Ergodic Theorem for Markov Chains, $\overline{X}_N$ converges almost surely to $\mu = E\{\pi(\eta| p_{\text{data}})\}$, as $N\rightarrow\infty$ \cite{robert2005monte}. The idea of Markov Chain Monte Carlo is to use $\overline{X}_N$ to estimate $\mu$.

Under certain conditions on the Markov Chain, a Central Limit Theorem (CLT) will assert the existence of an asymptotic variance $\gamma^2$ and the convergence in distribution
\[\sqrt{N}(\overline{X}_N-\mu)\overset{\mathcal{D}}{\longrightarrow} N(0, \gamma^2).\]

For example, one useful version of a CLT for Markov Chains is given in \cite{kipnis1986central}. It is a more sophisticated functional central limit theorem (see also \cite{geyer1992practical}),  so we state here a simplified version of it as in \cite{robert2005monte}.

\begin{theorem}[Central Limit Theorem] If $\{X_N\}$ is aperiodic, irreducible, and reversible with invariant distribution $\pi$, $g$ a functional in $L^1$, and if
\begin{align*}
0<\gamma_g^2 =& E_{\pi}\bigg\{\big(g(X_0)-E_{\pi}[g]\big)^2\bigg\}\\
&+2\sum_{k=1}^{\infty}E_{\pi}\bigg\{\big(g(X_0)-E_{\pi}[g]\big)\big(g(X_k)-E_{\pi}[g]\big)\bigg\} \\
&\quad < +\infty,
\end{align*}
then 
\[\frac{1}{\sqrt{N}}\left(\sum_{n=1}^N(g(X_n)-E_{\pi}[g])\right)\overset{L}{\longrightarrow}N(0, \gamma_g^2).\]
\end{theorem}

In particular, this is true for the Metropolis-Hastings method, since it produces a reversible Markov Chain \cite{robert2005monte}. 

From here, we can construct confidence intervals for $\overline{X}_N$ and estimate the minimum number of sample values $N$ required for $\overline{X}_N$ to attain a certain precision. Denote by $\widetilde{\gamma}^2$ an estimate of the asymptotic variance $\gamma^2$. 
By the above CLT, an approximate $1-\alpha$ confidence interval for the mean $\overline{X}_N$ is
\[\left(\overline{X}_N-z_{1-\alpha/2}\frac{\widetilde{\gamma}}
{\sqrt{N}},\quad \overline{X}_N+z_{1-\alpha/2}\frac{\widetilde
{\gamma}}{\sqrt{N}}\right).\]
Here, $z_{1-\alpha/2}$ is the quantile from either the standard Normal distribution or the $t-$student distribution. The standard deviation used in the above confidence interval is also called the \textit{Monte Carlo Standard Error} (MCSE) \cite{robert2005monte},
\[\text{MCSE} = \frac{\widetilde{\gamma}}{\sqrt{N}}.\]
It can be used as stopping criterion for the MCMC method \cite{vats2019multivariate}.  

To find the expression for $\gamma^2_g$ as in the CLT theorem, we start by calculating the variance of the estimator $\overline{X}_N$. The argument that follows is already standard and available in many references, for example \cite{sokal1997monte,priestley1981spectral}. Since the $X_i$ are no longer independent, we have to keep track of their covariances,
\begin{align*}
\text{Var}(\overline{X}_N) = \frac{1}{N^2}\bigg(&\sum_{t=1}^N\text{Var}(X_t)\\
&+2\sum_{s=1}^{N-1}\sum_{t=s+1}^N\text{Cov}(X_s, X_t)\bigg).
\end{align*}
We're interested in the asymptotic variance, that is, we want to see what happens when $N\rightarrow\infty$. For that, we use concepts from Time Series Analysis. If we assume that the Markov Chain $\{X_t\}$ has already converged, then it will be a second-order stationary process \cite{priestley1981spectral}. This allows us to claim that there exists an \textit{autocovariance function} $R$ such that
\[R(t-s)=\Cov(X_s, X_t), \qquad\text{for all }s, t.\]
This function is symmetric, i.e., $R(-k)=R(k)$, and its normalized version is called the \textit{autocorrelation function}, 
\[\rho(k) = \frac{R(k)}{R(0)}.\]

Now, 
\begin{align*}
\Var(\overline{X}_N) &= \frac{1}{N^2}\sum_{r,s=1}^N\Cov(X_r,X_s) \\
&= \frac{1}{N^2}\sum_{r,s=1}^NR(r-s) \\
&= \frac{1}{N}\sum_{k=-(N-1)}^{N-1}\left(1-\frac{|k|}{N}\right)R(k)
\end{align*}

Assuming that $R(k)$ is summable, that is, it satisfies $\sum_{k=0}^{\infty}R(k) < \infty$, then (see \cite{anderson2011statistical} for a rigorous proof using Cèsaro sums),
\begin{align*}
\Var(\overline{X}_N)&\approx \frac{1}{N}\left(\sum_{k=0}^{\infty}\frac{R(k)}{R(0)}\right)R(0) \\
&= \frac{1}{N}\left(\sum_{k=-\infty}^{\infty}\rho(k)\right)R(0) \\
&= \frac{1}{N}\left(\sum_{k=-\infty}^{\infty}\rho(k)\right)\sigma^2.
\end{align*}
The argument that $R(k)$ is indeed summable, at least in the case for finite dimensional state spaces, comes from a spectral radius argument \cite{green1992metropolis,sokal1997monte}.

The variance of $\overline{X}_N$ for large $N$ can then be approximated by 
\[\Var(\overline{X}_N)\approx\frac{\IACT}{N}\sigma^2,\]
where 
\[\IACT = \sum_{k=-\infty}^{\infty}\rho(k)=1+2\sum_{k=1}^{\infty}\rho(k)\]
is called the \textit{Integrated AutoCorrelation Time}.

We recall that, if $\{x_1, \ldots, x_N\}$ came from i.i.d. random variables, as is the case in regular Monte Carlo, then $\Var(\overline{X}_N)=\frac{1}{N}\sigma^2$. Comparing this expression with the variance we obtained for the MCMC sequence, we find an informal interpretation for the $\IACT$: it is the inefficiency factor accounting for the correlation between the MCMC sample values. 

It is also possible to use the concept of the \textit{spectral power density function} $f$ of a time series to write the expression for the IACT \cite{priestley1981spectral}. It will use the property that $f$ is the Fourier transform of the autocovariance function $R$. The IACT then assumes a simpler form, 
\[\IACT = 2\pi f(0).\]
Although the spectral power density function is not known in closed form for most of the MCMC methods, this idea is the key to fast algorithms to calculate the IACT based of the Fast Fourier Transform. This will be discussed in Section \ref{estimateESS}. 

Some works in Computational Physics also mention an \textit{Exponential Autocorrelation Time}. It is defined as the coefficient $\tau_{\text{exp}}$ such that $\rho(t)\approx e^{\frac{-t}{\tau_{\text{exp}}}}$ for large $t$ \cite{sokal1997monte,madras1988pivot}. This is related to the spectral radius of the transition matrix of the Markov Chain, in MCMC of finite state spaces. In those papers, the Integrated Autocorrelation Time, denoted here IACT* to distinguish it from ours, is defined as
\[\text{IACT*} = \frac{1}{2}\text{IACT} = \frac{1}{2}+\sum_{k=1}^{\infty}\rho(k),\]
to make it coincide with the exponential time in some circumstances. Since we're interested in the ESS and the interpretation that the IACT summarises the ``memory'' of our MCMC methods, we don't use such version here.  

Finally, the \textit{Effective Sample Size}, for a sample of size $N$, is defined as
\[\ESS(N) = \frac{N}{\IACT} = \frac{N}{1+2\sum_{k=1}^{\infty}\rho(k)}.\]
It leads to the informal interpretations that a sample of size $N$ from the MCMC is equivalent to an i.i.d. sample of size ESS to calculate the Monte Carlo expectation, and that IACT gives the time required to get one independent value from the MCMC. 

There aren't many theoretical results about the ESS itself, but only for the autocorrelation and autocovariance functions. For example, it is possible to prove that $R(k)\rightarrow 0$ for some MCMC methods \cite{robert2005monte}, and the following is also an interesting result by Geyer.

\begin{theorem}[\cite{geyer1992practical}] For a stationary, irreducible, reversible Markov Chain with autocovariances $\gamma_t$, let $\Gamma_m = \gamma_{2m}+\gamma_{2m+1}$ be the sums of adjacent pairs of autocovariances. Then $\Gamma_m$ is a strictly positive, strictly decreasing, strictly convex function of $m$.	\end{theorem}
	
It can be used to estimate the IACT, in what are called  ``initial sequence estimators'' \cite{geyer2011introduction}. The idea is to keep adding pairs of autocovariances until such pair becomes 0. However, if the Markov Chain is not reversible, it is advisable to use another method. 
 
\subsection{Example: the AR(1) process}\label{arprocess}


A simple example of a Markov Chain for which theoretical results are available is the first-order Autoregressive Process, or AR(1) \cite{kroese2013handbook,priestley1981spectral,robert2005monte}. It is also used by software packages as a test case (for example, the R package \textit{mcmcse}). 

An AR(1) process is defined by the relation
\[X_t=aX_{t-1}+\epsilon_t,\]
where $a$ is the parameter that specifies the model, and the $\epsilon_t\sim N(\mu_{\epsilon}, \sigma_{\epsilon})$ are assumed to be independent in $t$.

It can be shown that \cite{priestley1981spectral}  
	\[E[X_t] = \mu_{\epsilon}\left(\frac{1-a^t}{1-a}\right),\quad \sigma_{X_t}^2=\sigma_{\epsilon}^2\left(\frac{1-a^{2t}}{1-a^2}\right).\]
We can see that, when $|a|<1$, the Markov Chain $\{X_t\}$ converges to \[N\left(\frac{\mu_{\epsilon}}{1-a}, \frac{\sigma_{\epsilon}^2}{(1-a^2)}\right)\] as $t\rightarrow\infty$. It can also be shown that this is the stationary distribution for the chain \cite{robert2005monte}.

To simplify the example, let $\mu_{\epsilon}=0$. Then the unique stationary distribution of this Markov Chain is $N(0, \sigma_{\epsilon}/(1-a^2))$. 

It is possible to find a closed expression for its spectral power density \cite{priestley1981spectral},
\[f(\omega) = \frac{1-a^2}{2\pi(1-2a\cos\omega +a^2)}, \quad -\pi\leq\omega\leq\pi.\]
In particular, 
\[f(0)=\frac{1+a}{2\pi(1-a)}.\]
	
This will already give us its IACT, but this model is simple enough so that we can estimate it as the infinite sum given in the previous section. The autocorrelation function for this chain is \cite{priestley1981spectral},
\[\rho(r)=a^{|r|},\quad r\in\mathcal{Z},\]
so the infinite sum of the IACT becomes again
\[\text{IACT}=1+2\sum_{k=0}^{\infty}\rho(k)=\frac{1+a}{1-a}.\]
If we need an example with a specific IACT, we can find the required $a$ by solving the above equation. 
	
Finally, the variance for the Monte Carlo estimator becomes
\[\Var(\overline{X}_N)=\frac{1}{N}\frac{1}{(1-a)^2}\sigma_{\epsilon}^2.\]

\section{Estimating the ESS}\label{estimateESS}

By the discussion from the previous section, we notice that estimating the asymptotic variance of an MCMC method, its IACT, or its ESS (given a number $N$ of iterations so far) are all related problems. 

Except for the Batch Means approach, which directly estimates the asymptotic variance, every other method first calculates the IACT and then uses it to find the ESS. We can either use parametric methods and fit the output into a model with known IACT, like the AR(p) fitting method, or we can first estimate the autocovariance function $R$ and then estimate the IACT's infinite sum. 

We will experiment on the consistency of the ESS using AR(1) models, but we can understand some simple theory by writing it as a function of the consistency of the IACT. By its definition, 
\[\ESS = \frac{N}{\IACT}.\]
Using the approximation \cite{casella2002statistical}
\[\Var\left(\frac{1}{X}\right) \approx \left(\frac{1}{E(X)}\right)^4\Var(X),\]
we have that, assuming that the estimators of the IACT are unbiased, 
\[\Var(\widehat{\ESS}) = \Var\left(\frac{N}{\widehat{\IACT}}\right)\approx N^2\frac{\Var(\widehat{\IACT})}{\IACT^4}\]

This indicates that, if the variance of the estimator of the IACT does not decrease at least in quadratic order, than we have no hopes that the estimator of the ESS will be consistent. For example, assume that $\Var(\widehat{\IACT}) \approx \frac{C}{N}$. This is likely to be a very optimistic assumption, and in our numerical experiments in Section \ref{arexperiments} we found this decrease to be much slower. Then, 
\[\Var(\widehat{\ESS}) \approx\frac{N^2}{\IACT^4}\times\frac{C}{N} = N\frac{C}{\IACT^4}.\]
Thus, in this case the variance of the ESS estimate will grow proportionally to N.

In this section, we go over some methods to estimate the ESS. We start with the estimation of the autocovariance function $R$, and then the ESS-Bulk and spectral window estimators that use that function. We then discuss the Batch Means and AR(p) fitting methods. 

\subsection{Estimates for  $R(k)$}\label{rksec}

The material in this section was adapted from the classical textbooks in Spectral Time Series Analysis, \cite{priestley1981spectral,anderson2011statistical,percival1993spectral}.

We first recall the definition of the autocovariance function $R$,
\[R(k) = \Cov(X_{t_0}, X_{{t_0}+k})=E\{(X_{t_0}-\mu)(X_{{t_0}+k}-\mu)\},\]
and this should be the same value for any $t_0$, since the chain is assumed to already be stationary (see discussion in Section \ref{MCMCtheo}).

A simple and intuitive estimator for $R(k)$ is defined, for $k$ positive, as
\[\widehat{R}(k) = \frac{1}{N}\sum_{i=1}^{N-k}(x_i-\overline{X}_N)(x_{i+k}-\overline{X}_N).\]
Since R is symmetric, this also gives the values for negative $k$. The denominator of $N-k$ would be more intuitive than $N$, but the formula presented here has better statistical properties \cite{priestley1981spectral}. Also, we notice that the value of $R(k)$ can only be estimated for $k$ up to $N-1$, since there will be no pair of observations with larger lag than that.  

There is an intuitive argument for why the estimates for $R(k)$ get worse as $k$ becomes larger. This is because there will be fewer and fewer pairs of values $(x_t, x_{t+k})$ for the average above (precisely $N-|k|$ such pairs). But for an expectation on the stationary distribution of the Markov Chain, a reliable estimate would require sufficient iterations that go through the entire support of the distribution.

It is possible to obtain some theoretical properties on $\widehat{R}$, and they are summarised in \cite{priestley1981spectral}. For example, 
\begin{align*}
E[\widehat{R}(k)] &= R(k) - \frac{|k|}{N}R(k) -\Var({\overline{X}_N}) \\
&= R(k) - \frac{|k|}{N}R(k) -\frac{2\pi\sigma^2(N-|k|)}{N^2}f(0).
\end{align*}
From this formula, we can justify a property observed in practice: as $k\rightarrow N$, the second term will almost entirely cancel the first, and the estimate will be close to zero. This means that the estimated autocorrelation is expected to go to zero faster than it should, if N is not large enough. 

When the mean $\mu$ is known, we can ignore the term with $\Var({\overline{X}_N})$, and the bias is simple,
\[\text{bias}(\widehat{R}(k)) = \frac{|k|}{N}R(k).\]

For a more general case that includes the errors when estimating the mean, there are some theoretical results available $\widehat{R}(k)$, but they are either too complicated to be interpreted or require strong simplifying assumptions. For example, the variance of the estimator is given in \cite{priestley1981spectral}, but it requires fourth-order stationarity of the chain and $\mu$ still has to be known in advance. The textbook \cite{anderson2011statistical} also provides quite many expectation and variance formulas, but they are just as difficult to interpret. Still, for a fixed lag $k$, it can be said that the order of the bias for $\widehat{R}(k)$ is of order $O(1/N)$, and in many cases, the variance is also $O(1/N)$. 

We notice that the formula for the IACT uses $\rho(k)=R(k)/R(0)$, and not $R(k)$. However, its properties are even more complicated, and also require strong assumptions on the chain \cite{priestley1981spectral}.

Most software packages rely on the Fast Fourier Transform to calculate the $\widehat{R}(k)$. This is based on a relation between $\widehat{R}(k)$ and an estimate of the spectral power density $f(x)$ called \textit{periodogram}. We start by centering the sample $\{x_1, \ldots, x_n\}$ by subtracting its sample mean. This defines a new sequence $\{\widetilde{x}_1, \ldots, \widetilde{x}_N\}$ of mean zero but same autocorrelation as the original one. From there, the periodogram estimator is defined as
\[I_N(\omega)=\frac{1}{N}\left|\sum_{t=1}^N\widetilde{x}_te^{-i\omega t}\right|^2,\]
which can be calculated rather quickly, by means of a Fast Fourier Transform, then taking its modulus squared. We then return to $\widehat{R}(k)$ by doing an inverse Discrete Fourier Transform \cite{priestley1981spectral}. Details can be found in \cite{priestley1981spectral} and \cite{percival1993spectral}. In some works, the periodogram formula above is different up to a multiplying constant. 
Surprisingly, this procedure using the FFT is equivalent to calculating $\widehat{R}(k)$ by the formula in the beginning of this section, but faster.

It's worth mentioning that an important reference \cite{gelman2013bayesian} estimates the autocovariance function $R$ based on another statistic called \textit{variogram}. Lately, however, the FFT method has been preferred \cite{vehtari2021rank}. 

Once the autocorrelation function has been estimated, the task becomes to find a way to truncate the infinite sum that defines the IACT. If the chain is known to be reversible, then we can use Geyer's Theorem from Section \ref{MCMCtheo}. Otherwise, we need another approach, like one of the methods below. 

\subsection{Spectral window methods}\label{window}


After obtaining $\widehat{R}(k)$, we can estimate the autocorrelation function $\widehat{\rho}(k)$, from $k=0$ to $k=N-1$. We recall that
\[\text{IACT} = \sum_{k=-\infty}^{\infty}\rho(k).\]

The idea of a \textit{spectral window} is to estimate the IACT by the weighted sum
\[\widehat{\text{IACT}} = \sum_{k=-(N-1)}^{k=(N-1)}\lambda(k)\widehat{\rho}(k).\]
The sequence $\{\lambda(k)\}$ is called a \textit{lag window}, and its discrete Fourier transform is the \textit{spectral window}. This idea first came up as a way to estimate the spectral density function $f$ of a time series \cite{priestley1981spectral}. Since $\text{IACT}=2\pi f(0)$, it can also be applied to estimate the IACT and ESS. 

There are many choices for the function $\lambda$ of weights, but we mention here only three. Textbooks of Spectral Series Analysis have extensive lists of attempted weight functions and some of their properties \cite{priestley1981spectral} \cite{anderson2011statistical} \cite{percival1993spectral}. 

The simplest window is just a truncation of the sum up to the lag $M$, 
\[\lambda(k) = \left\{\begin{array}{l l}
1,&\text{ if }|k|\leq M,\\
0,&\text{ if }|k|> M,\end{array}\right.,\]
for which some bounds on the estimation error are available in \cite{madras1988pivot} and \cite{wolff2004monte}.

The more sophisticated ones below are the Bartlett Window,
\[\lambda(k) = \left\{\begin{array}{l l}
 1-|k|/M,&  \text{if }|k|\leq M,\\
 0,& \text{if }|k|>M,\end{array}\right.\]
and the Tukey Window,
\[\lambda(k) = \left\{\begin{array}{l l}
1-2a+2a\cos(\pi kM),& \text{if }|k|\leq M,\\ 
0,& \text{if }|k|>M,\end{array}\right.\] 
which has an additional parameter $0<a\leq 1/4$.

These last two methods are implemented in the \textit{mcmcse} package in R. We refer to \cite{priestley1981spectral} for the intuition and analysis of those methods.

\subsection{The ESS-Bulk}\label{essb}


We present here the method to calculate the ESS that was proposed in \cite{vehtari2021rank}, which is currently the default estimator used by the popular software packages \textit{STAN} and \textit{ArviZ}. 
It is a distinct approach from the previous methods in two ways. Firstly, this new approach combines the ESS with the Potential Scale Reduction Factor (PSRF) into just one value that will summarize the quality of the estimate of the MCMC. Secondly, it preprocesses the chain through a rank normalization transformation.

To understand the formula for the method, we provide a quick summary of the PSRF. It was first introduced in \cite{gelman1992inference}, but its popular modern version is presented in \cite{gelman2013bayesian}, and a newer version is in \cite{vehtari2021rank}. 

The question that the PSRF intends to answer is: given M chains (j-index) simulated independently in parallel, have they run for long enough so that they are sampling from the same distribution? If yes, then we have evidence that those chains have converged to their limiting distribution. The PSRF will be a number $\widehat{R}^2$ such that, if close enough to 1, will indicate that this is indeed happening. However, how close to 1 we should get is a problem in itself. For example, \cite{vats2021revisiting} have suveyed some papers and found that this choice is rather arbitrary. More crucially, a PSRF of 1 is a necessary, but not sufficient condition for the chains to have converged. 

If fact, modifications have been proposed so that a PSRF of 1 has more chances of indicating true convergence. Typical implementations of the PSRF drop the first half of each chain as burn-in and then estimate the quantity on the second half. This is already not good in our setting, since slow-mixing chains have to be run for longer, and dropping half the chain is too expensive. But the modern modification is even worse: after removing the first half of the chain, the new method splits the remaining chains in two, so we end up having $2M$ chains with $N/4$ values each. Thus, the PSRF will be an overestimate, and a convergence criterion based on it will force the chains to run for way longer than necessary. 

The formula for PSRF resembles a test statistic of Analysis of Variance, in that we compare some variances to check if the averages are close enough. Assume each of the $M$ independent chains (j-index) have N values (i-index) $X_{ij}$ each.  Recall that we can have two types of averages: the complete average $X_{\cdot\cdot}$, taking into account all the simulated values from all parallel chains, and the average $X_{\cdot j}$ for each of the chains that were run in parallel. 
\[X_{\cdot j} := \frac{1}{N}\sum_{i=1}^NX_{ij},\quad X_{\cdot\cdot} := \frac{1}{NM}\sum_{j=1}^M\sum_{i=1}^N X_{ij}\]

Likewise, we'll have two types of variances. One is the variance of the average of each chain against the overal average, called \textit{between-sequence variance},
\[\displaystyle{B := \frac{N}{M-1}\sum_{j=1}^M(X_{\cdot j}-X_{\cdot\cdot})},\]
and the second is the usual variance of each chain taken as an isolated sample, called \textit{within-sequence variance}. It is given by 
\[W := \frac{1}{M}\sum_{j=1}^Ms^2_j,\quad\text{with}\quad s^2_j:=\sum_{i=1}^N(X_{ij}-X_{.j})^2.\]

We still have another variance, $\widehat{var}$, which is an estimate of the marginal posterior variance. It can be calculated using a linear combination of the previous variances. Finally, the PSRF, also commonly denoted by $\widehat{R}$, is the square root of the ratio
\[\widehat{R}^2 = \frac{\widehat{var}}{W} = \frac{\frac{N-1}{N}W+\frac{1}{N}B}{W}.\]

Now, we can start discussing how the ESS-Bulk is calculated. The name ``Bulk'' comes from a preprocessing step that is done by the method. As an attempt to make the estimator generalizable to distributions that are not symmetric or Gaussian, rank-normalization is applied. First, each sample value $x^{(i)}$ of the chain is transformed into its rank-version $r^{(i)}$, that is, $1, 2, 3,\ldots,$ and so on.  Then, the Gaussian CDF $\Phi$ is used to obtain the rank-normalized output $z^{(i)}$,
 \[z^{(i)} = \Phi^{-1}\left(\frac{r^{(i)}-\frac{3}{8}}{S-\frac{1}{8}}\right).\]
As it will be observed in the numerical experiments, this does not seem to be a good idea when the chains are too autocorrelated. Even if the posterior is Gaussian, it will take a while for it to be truly symmetric, so the output will be distorted. When the chain is small, this will produce a large variance on the IACT estimator. 
 
Once rank-normalization and the PSRF are calculated, we estimate the autocorrelations $\rho_m(t)$ of each individual chain by traditional means. In the software packages that were mentioned, this is found by the FFT method explained in Section \ref{rksec}. Then, the autocorrelations $\widehat{\rho}$ that will be used to estimate the ESS bulk are defined as
\[\widehat{\rho}(t) := 1-\frac{1}{\widehat{R}}+\frac{\frac{1}{M}\sum_{m=1}^Ms_m^2\rho_m(t)}{\widehat{var}}.\]
	
The idea is that, if the Markov Chains converged, then $\widehat{R}\approx 1$ and the variances $s_m $ and $\widehat{var}$ coincide for all chains $m$. This implies that $\widehat{\rho}(t) = \rho_m(t)$, that is, the autocorrelation estimates will be the same as $\rho_m(t)$. 
	
The other limiting case is when $\widehat{R}$ is much larger than $1$. Looking at the expression for $\widehat{\rho}(t)$, we see that the last fraction resembles $\frac{1}{\widehat{R}^2}$, only that the variances that make up $W$ are now weighted by $\rho_m$. Since $|\rho_m| < 1$, then this fraction is less than $\frac{1}{\widehat{R}^2}$. For large $\widehat{R}$, the term will be very small and we end up with $\widehat{\rho}(t) \approx 1 - \frac{1}{\widehat{R}}$. 

In our view, such $\widehat{\rho}$ makes no sense. As discussed in \textit{Section \ref{MCMCtheo}}, the theory of the ESS  requires two things: First, that the autocorrelations be summable, which may no longer the case here; but second, that the chain has already converged in distribution. That is, if the chains have not converged yet, they are not a stationary time series and the formula for the ESS is no longer valid. That is, the PSRF is a prior and separate step to the estimation of the ESS. 

It is hoped that, in between those limiting cases, the inclusion of the PSRF in the autocorrelations will penalize somehow the lack of convergence of the chains. However, no further study was done on how this happens and what we are to expect. In our view, complicating the formula for the ESS only makes a difficult problem worse.

Once these $\widehat{\rho}$ are estimated, the methods calculate the IACT by one of the methods we previously mentioned. For example, STANs documentation says it uses Geyer's Initial Monotone Sum (see the end of \textit{Section} \ref{MCMCtheo}). It is important to mention that this is theoretically valid only for reversible chains, but the software package uses it anyway.

In a way, the ESS-Bulk only has chances of giving reliable estimates when it is large. The authors indicate that an ESS of 400 would give sufficient evidence of convergence. However, as it will be shown in the simulations, this method overestimates the IACT in such a way that, even when $\widehat{R}$ is not so large, say $\widehat{R}\approx 1.2$, the ESS remains stubbornly low for very long. Relying on this quantity to continue the MCMC sampler would make it run for much longer than needed, just to reach such ESS number of 400.  

Finally, distinct software packages may implement these ideas differently. For example, we used \textit{rstan}'s version of the ESS-Bulk, and some details were only available by reading the actual code, and not the documentation. There, no burn-in is performed on the chains given as input, and they are first split in half. This doubles the number of chains, but they now have only half their original length. Then, they are rank-normalized and fed into the ESS-Bulk method. So, even the PRSF is calculated after rank-normalizion.

\subsection{Batch means}


The method of \textit{batch means} is another popular method to estimate the asymptotic variance of the MCMC output, mainly for its simplicity: we do not have to go through the infinite sum of autocorrelations \cite{geyer2011introduction}. It is also called \textit{binning method} in \cite{wolff2004monte}. 

The original batch-means, now called \textit{non-overlapping} batch means, divides the $N$ samples into $k$ non-overlapping and adjacent batches of size $m$. That is, the output $\{x_1, x_2, \ldots, x_N\}$ is divided into $\{x_1, x_2, \ldots, x_m\}$, $\{x_{m+1}, x_{m+2}, \ldots, x_{2m}\}$, and so on. 

For each batch $j$, we can calculate its sample mean $\widehat{\mu}_{j}$. Now, suppose that those batches are large enough so that each $\widehat{\mu}_j$ is independent of the $\widehat{\mu}_l$ of a different batch. What we have is an approximately independent sample for the sample means, $\{\widehat{\mu}_1, \widehat{\mu}_2, \ldots, \widehat{\mu}_k\}$. Its variance will be precisely the variance of the CLT that we're trying to estimate, adjusted for the sample sizes being the batch size $m$.

So, we calculate
\[\Gamma_{bm}=\frac{1}{k}\sum_{j=1}^k(\widehat{\mu}_j-\widehat{\mu})^2\]
as an estimate for $\gamma^2/m$.

Originally, the number of batches was fixed, regardless of the increase of $n$. But further results in its theory have shown that such an estimator has poor asymptotic properties. For example, for the estimator to be consistent, it must have more and more batches as $n\rightarrow\infty$. Also, as correlation increases, large batch sizes are required \cite{damerdji1994strong}, since it is hoped that the batches are sufficiently large to be uncorrelated to one another. 

The issue we have with this method is then that it is more suitable for methods where convergence is faster. In subsurface simulations, the iterations are computationally expensive and the correlation is high. In summary, the chains will be too short for this method to produce good estimates. 

Flegal and Jones \cite{flegal2010batch} have tested different batch means parameters on AR(1) sequences. They mention the choice of number of batches for a fixed $N$ to be $N^{1/3}$ for not so high correlations, but $N^{2/3}$ for high ones. Moreover, the OBM has smaller variability than the BM method. 

The Overlapping Batch Means (OBM) is a modification of the original method, and it is meant to have better properties. Now, we allow the batches to overlap, effectively creating even more batches. It is a more computationally expensive method, and it has interestingly related to the Bartlett window method mentioned in Section~\ref{window} \cite{damerdji1994strong}.

As for the decrease in variance, we notice that even if $\widehat{\gamma}^2$ is consistent, this only means that $\frac{\text{IACT}}{\sqrt{N}}\rightarrow 0$, which is not of quadratic order. 

\subsection{AR(p) fitting}


The default method to calculate the ESS in R is by first fitting the sequence of values $\{x_1, x_2,\ldots,x_N\}$ into an Autoregressive Process of order p. Then, its IACT will be also calculated by its spectral power density $f$ evaluated at $0$. It is also the method implemented in the CODA package for R, mentioned in \cite{gamerman2006markov, robert2005monte} and used by the BUGS software \cite{lunn2000winbugs}.

Building on the AR(1) example, the Autoregressive Process of order $p$, denoted AR(p), is given by
\[X_t+a_1X_{t-1}+\ldots+a_kX_{t-p} = \epsilon_t,\] 
with parameters $a_1$, $a_2$, \ldots $a_k$. The random process $\epsilon_t$ is as in the AR(1) process. The study of the autocorrelation function will depend on the study of the process's characteristic polynomial, 
\[g(z) = z^k+a_1z^{k-1}+\ldots+ a_k.\] 
Details on how to fit a series into an AR(p) process and the estimation of its $f(0)$ can be seen in \cite{priestley1981spectral}. 

The disadvantage of this method is that AR(p) processes, for order $p$ larger than 1, are no longer Markov Chains, even though they can still be stationary processes. It may therefore not make sense to use this model as an approximation of our output. 

\section{Numerical results}\label{simulations}
	
\subsection{Computational environment}

The output analysis and estimation of the ESS and IACT were performed on a laptop with 8GB of RAM, an 8th-Generation intel core i5 processor, and a Linux operating system. We used RStudio version 2023.06.1 Build 524 for Linux, with R version 4.1.0. The R packages used were \textit{coda} version 0.19-4, \textit{mcmcse} version 1.5-0, and \textit{rstan} version 2.21.5.

The Markov chains were simulated on a GPU cluster running on an Intel Xeon E5-1620, with 32GB of RAM. Each chain was assigned to a different Nvidia Tesla K80 GPU with 12GB of dedicated RAM.  	
	
\subsection{Experiments with AR(1) models}\label{arexperiments}

\begin{figure*}[t]
\includegraphics[scale=\scaleval]{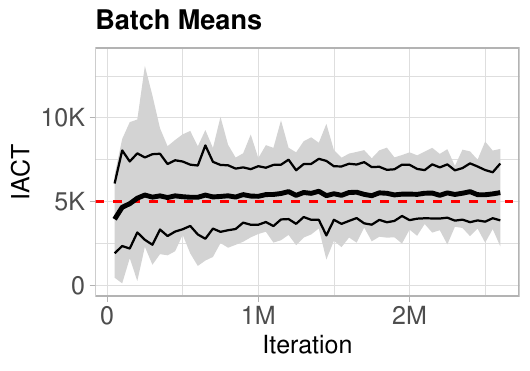}
\hspace{1cm}
\includegraphics[scale=\scaleval]{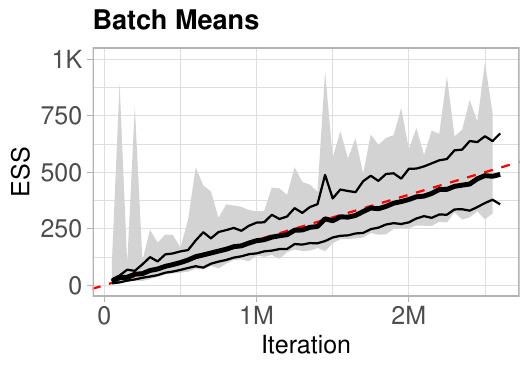}

\includegraphics[scale=\scaleval]{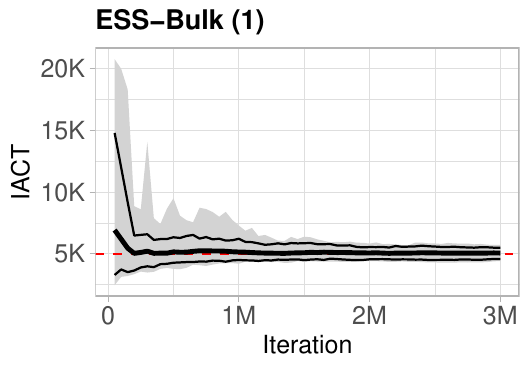}
\hspace{1cm}
\includegraphics[scale=\scaleval]{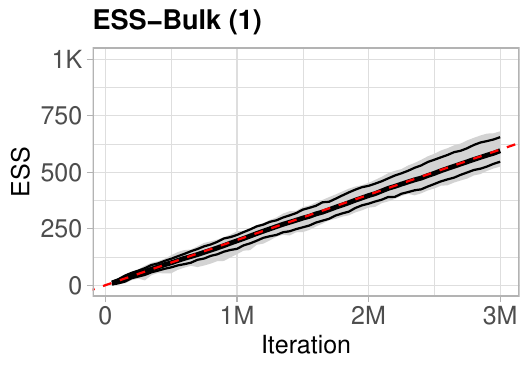}

\includegraphics[scale=\scaleval]{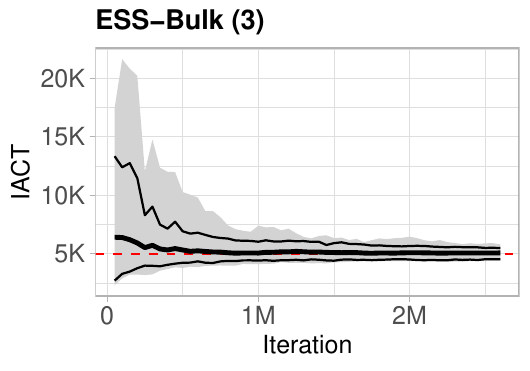}
\hspace{1cm}
\includegraphics[scale=\scaleval]{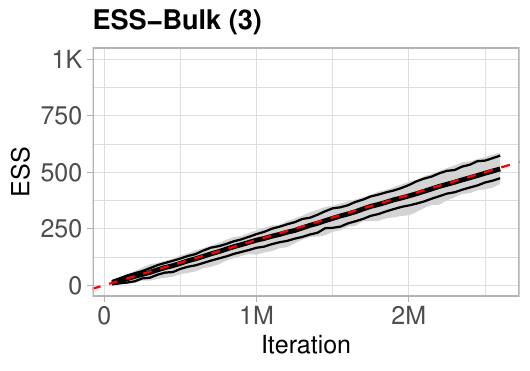}
\caption{Plots for the empirical distributions of some estimators for the IACT and ESS, using 100 independent AR(1) chains with IACT$=$5,000. The dashed line in red is the true value, and the area in gray is the full range of observed values. The lines in black are the 5\% percentile, mean and 95\% percentile. The first plot has a different scale than the others, to preserve its details.}\label{plotsar5k}
\end{figure*}

\begin{table*}[t]
\small
\caption{Results for AR(1) chains with IACT$=$5,000}\label{tar5k}%
\begin{tabular}{l|r|r|r|r|r|r|r|r}
\hline
 & \multicolumn{2}{|c|}{IACT at} &\multicolumn{2}{|c}{IACT at}&\multicolumn{2}{|c|}{ESS at} &\multicolumn{2}{|c}{ESS at}\\
 & \multicolumn{2}{|c|}{1.6M iterations} &\multicolumn{2}{|c}{2.6M iterations} & \multicolumn{2}{|c|}{1.6M iterations} &\multicolumn{2}{|c}{2.6M iterations}\\
 \hline
Method & Mean & SD & Mean & SD& Mean & SD & Mean & SD\\
\hline
AR(p) Fitting   & 4983.74 & 275.82  & 4982.04 & 223.10  & 322.02 & 17.81 & 522.91 & 23.42 \\
Batch means     & 5535.55 & 1057.40 & 5530.43 & 1040.25 & 300.95 & 65.67 & 490.19 & 116.28\\
Bartlett window & 5368.58 & 873.41  & 5319.21 & 835.88  & 306.54 & 54.18 & 502.60 & 92.27 \\
Tukey window    & 5730.14 & 918.37  & 5629.23 & 902.11  & 286.97 & 50.01 & 475.77 & 91.77 \\
ESS-Bulk 2      & 5131.08 & 778.05  & 5088.18 & 597.88  & 318.32 & 44.31 & 517.58 & 57.24 \\
ESS-Bulk 3      & 5106.89 & 386.81  & 5072.62 & 290.89  & 315.06 & 23.51 & 514.24 & 29.77 \\
\hline
ESS-Bulk 1$^*$  & 5062.57 & 307.32  & 5056.76 & 261.43  & 396.50 & 24.16 & 594.86 & 31.22 \\
\hline
\multicolumn{9}{l}{* No burn-in removed. Values are for iterations 2M and 3M, respectively}\\
\end{tabular}
\end{table*}

\begin{figure*}[t]
\includegraphics[scale=\scaleval]{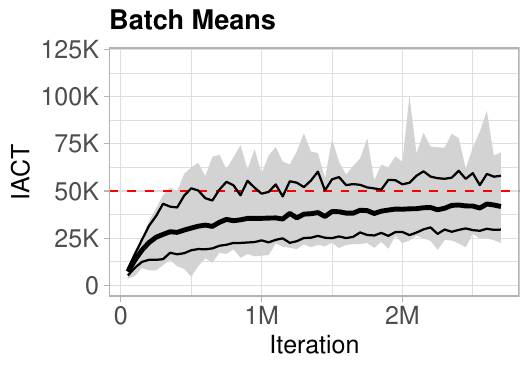}
\hspace{1cm}
\includegraphics[scale=\scaleval]{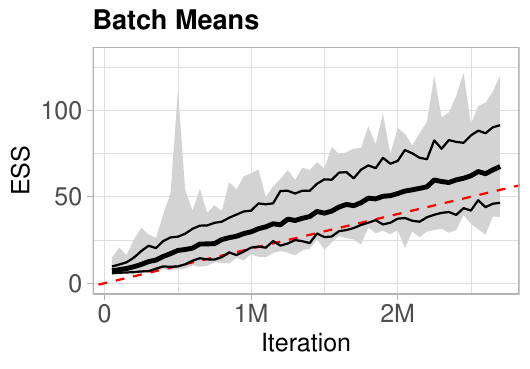}

\includegraphics[scale=\scaleval]{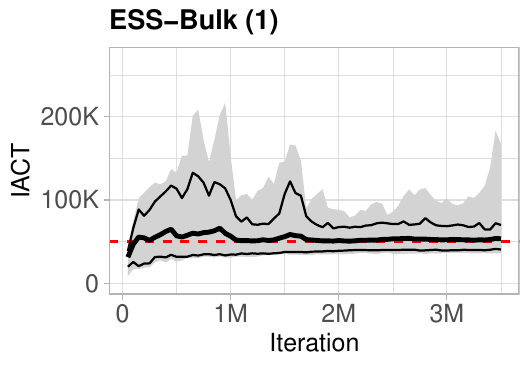}
\hspace{1cm}
\includegraphics[scale=\scaleval]{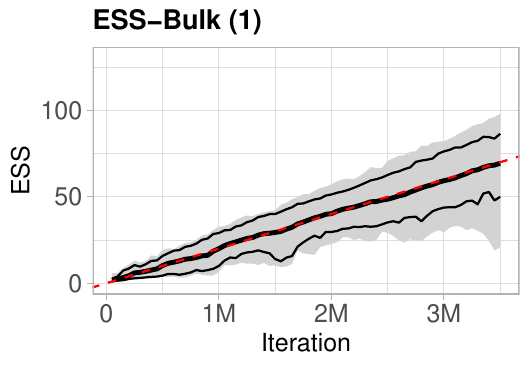}

\includegraphics[scale=\scaleval]{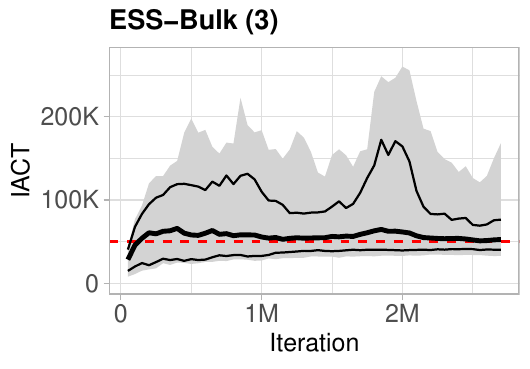}
\hspace{1cm}
\includegraphics[scale=\scaleval]{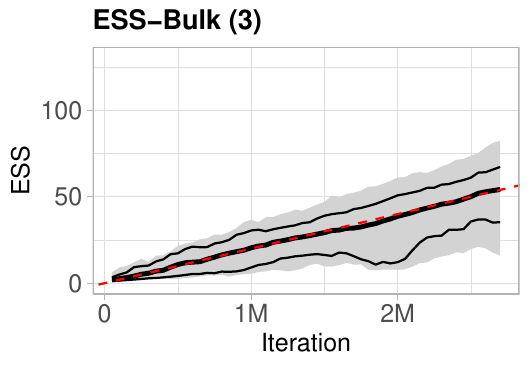}
\caption{Plots for the empirical distributions of some estimators for the IACT and ESS, using 100 independent AR(1) chains with IACT$=$50,000. The dashed line in red is the true value, and the area in gray is the full range of observed values. The lines in black are the 5\% percentile, mean and 95\% percentile. The first plot has a different scale than the others, to preserve its details.}\label{plotsar50k}
\end{figure*}

\begin{table*}[t]
\small
\caption{Results for AR(1) chains with IACT$=$50,000}\label{tar50k}
\begin{tabular}{l|r|r|r|r|r|r|r|r}
\hline
& \multicolumn{2}{|c|}{IACT at} &\multicolumn{2}{|c}{IACT at}& \multicolumn{2}{|c|}{ESS at} &\multicolumn{2}{|c}{ESS at}\\
& \multicolumn{2}{|c|}{iteration 1.7M} &\multicolumn{2}{|c}{iteration 2.7M}& \multicolumn{2}{|c|}{iteration 1.7M} &\multicolumn{2}{|c}{iteration 2.7M}\\
\hline
Method & Mean & SD & Mean & SD& Mean & SD & Mean & SD\\
\hline
AR(p) Fitting   & 47394.95 & 8026.07  & 48364.37 & 6935.97  & 36.94 & 6.53  & 57.01 & 8.44 \\
Batch means     & 39793.21 & 8913.74  & 41774.31 & 8915.26  & 44.86 & 10.09 & 67.52 & 14.26\\
Bartlett window & 39982.08 & 7909.22  & 42535.51 & 7927.86  & 44.13 & 8.52  & 65.62 & 11.89\\
Tukey window    & 42677.23 & 8532.05  & 45368.46 & 8529.56  & 41.41 & 8.21  & 61.58 & 11.38\\
ESS-Bulk 2      & 65437.50 & 62959.25 & 57954.46 & 47962.56 & 35.91 & 13.72 & 56.13 & 17.31\\
ESS-Bulk 3      & 58672.99 & 22058.89 & 52524.43 & 16548.19 & 31.86 & 8.43  & 54.23 & 10.26\\
\hline
ESS-Bulk 1$^*$  & 53335.01 & 10556.46 & 53372.58 & 17920.51 & 48.60 & 9.06 & 69.40 & 12.91\\
\hline
\multicolumn{9}{l}{* No burn-in removed. Values are for iterations 2.5M and 3.5M, respectively}\\
\end{tabular}
\end{table*}

In \textit{Section \ref{rksec}}, we have mentioned some theoretical results for the estimators of the autocovariance and autocorrelation functions. Since their deduction required many simplifying assumptions, those results may no longer be valid for complicated Markov Chains that simulate inverse problems.

To have an idea of what the sampling distributions of the estimators can be, we used simulations of the AR(1) Markov Chain described in \textit{Section~\ref{arprocess}}. The idea to use this example is far from new, and one can see it for example in \cite{emcee, flegal2010batch}. There also exists an argument against using AR(1) chains as a starting point: The claim is that they are too simple an example and therefore our conclusions would be of limited usefulness \cite{geyer2011introduction}. However, we will see that the estimators for the ESS have significant limitations even for this model, thus making these results worthwhile. For example, the very important observation that the \textbf{ESS estimators may not in practice be statistically consistent has never been made before}. It is, though, an easy conclusion from these experiments.   

Since we know that the IACT in inverse problems can be quite large, we started by investigating what happens to those estimators as the IACT is 5000. For that, we ran 100 independent AR(1) chains with the parameter $a$ corresponding to the desired IACT (\textit{Section~\ref{arprocess}}). Each independent chain was run up to 3 million iterations, from which we removed the initial 400k iterations as burn-in. The methods to calculate the ESS were then applied to the remaining 2.6M iterations of each chain. Table \ref{tar5k} reports the IACT, ESS and their respective standard deviations after 1.6M, and then 2.6M, iterations. Some plots were grouped in \textit{Figure} \ref{plotsar5k}. The remaining plots are available in the Appendix. 

In \textit{Section \ref{essb}}, we explained that the ESS-Bulk method incorporates the PSRF into the estimation of the ESS. However, our chains are highly autocorrelated, so a burn-in will only make the PSRF calculation worse by forcing it to start over and thus being an overstimate. Therefore, our first idea to use the ESS-Bulk estimator was to not remove any sample values as burn-in. We labeled this approach as \textit{ESS-Bulk 1}. To be compatible with how this method will be used in the inverse problem, we grouped those chains into 4 chains each. The way we did it was with a ``bootstrap" idea: we sampled 4 numbers ranging from 1 to 100 without replacement. Repeating this 100 times, we have 100 experiments with some combination of 4 independent chains from the original batch. Also, because no sample values were removed as burn-in, in Table \ref{tar5k} the estimates for its IACT and ESS are actually for 2M and 3M iterations. 

We then performed two more variations using the ESS-Bulk estimator. For those, the same amount of initial sample values was removed as burn-in as for the other methods. For the variation \textit{ESS-Bulk 2}, we fed the estimator one chain at a time, as an attempt to bypass the use of the PSRF. However, as explained in Section \ref{essb}, the code from \textit{rstan} will automatically split the chain in 2, so the PSRF will still be used even if just one chain were used as input. The last variation was named \textit{ESS-Bulk 3}, and it is how the \textit{rstan} documentation recommends the code to be used. We fed the same list of 4 by 4 chains as in \textit{ESS-Bulk 1}, with the difference that the 400k initial iterations were removed as burn-in.

The data in Table \ref{tar5k} shows us that \textbf{the standard deviation decreases very slowly in the estimation of the IACT}, and it remained roughly of the same order of magnitude as the estimate itself, for millions of iterations. As a first consequence, the standard deviation of the ESS increases as the number of iteration grows, and \textbf{this is what we mean when we claim that those estimators are not consistent}. Secondly, the estimate of the IACT can then only be trusted up to its order in magnitude, if we consider two standard deviations above and below the mean. For example, if the estimated IACT is 5k, then it may actually be anywhere between 3k and 7k. This matters if we want to use this quantity to compare two MCMC methods. 

Now, to understand what happens if the IACT is increased, a new batch of experiments was performed with 100 independent AR(1) chains whose IACT is 50,000. The number of iterations was increased to 3.5 million for each chain, of which the initial 800 thousand were removed as burn-in. This leaves us with 2.7M iterations, and a similar \textit{Table \ref{tar50k}} summarizes the results for iterations 1.7M and 2.7M, respectively, after the burn-in. Some plots for those experiments were grouped in \textit{Figure \ref{plotsar50k}}, and the complete set of plots is given in the Appendix. 

From Table \ref{tar50k}, we notice that even after 2.7M iterations, all methods apart from the ESS-Bulk still slightly underestimated the IACT on average. Moreover, they still have a standard deviation that is roughly the same after 1M iterations, and that are of the same order of magnitude as the estimate of IACT. This is the same conclusion as in the initial experiments with IACT of 5000. The most notable difference is in the estimation with the ESS-Bulk methods. They performed worse in terms of variance, and this large variation is also present in the plots of \textit{Figure \ref{plotsar50k}}. So, for a larger IACT, the ESS-Bulk starts to misbehave, as we predicted in \textit{Section \ref{essb}}. Also, since the limiting distribution of this example is Gaussian, the rank normalization step is not expected to distort much of the results, although there will still be some penalization while the number $N$ of iterations is not large enough. This effect may be worse for other limiting distributions, especially if they are not symmetric. 

Lastly, we notice that the AR(p) Fitting method outperformed every other method. This was expected, since this example is an AR(p) sequence itself. 

\subsection{Experiments for the inverse elliptic problem}\label{mcmcmethods} 

We now turn to the simulation of the posterior distribution for the inverse problem introduced in Section \ref{elliptic}. Two different MCMC methods were used to simulate this problem, and our goal was to analyze and compare their outputs using the ESS.  

The first MCMC method is the Two-Stage Delayed-Acceptance method \cite{christen2005markov}. Its main idea is to calculate the likelihood of the proposal on a coarse grid first, and perform an intermediate acceptance step. Only if this new sample is accepted do we calculate the likelihood on the fine grid, and a final acceptance step is then performed using this more accurate simuation. We used the code implemented by \cite{ali2021multiscale}, that uses a preconditioned Crank-Nicolson step instead of a random walk as a proposal distribution \cite{cotter2013mcmc}. 

The second MCMC method is the Multiscale Sampler \cite{ali2024multiscale}. It builds upon the Delayed-Acceptance MCMC and introduces the idea of decomposing the domain of the problem and performing MCMC locally. It uses local sampling and local averaging, for a non overlapping domain decomposition. Its details are more sophisticated, and they can be checked in the reference given. 

The most computationally expensive part of the MCMC algorithms for inverse problems is the solution of the forward problem. At each MCMC iteration, a differential equation has to be solved. Both MCMC methods were coded to use the same elliptic differential equation solver to be run on a GPU. It is a code-optimized conjugate gradient preconditioned by the algebraic multigrid method called Parallel Toolbox (\cite{liebmann2006user,neic2012algebraic}). In essence, the discretization is a 5-point cell-centered finite difference method, which is known to be equivalent to the finite element method with Raviart-Thomas of order 0 \cite{douglas1997numerical}. 

\begin{table*}[t]
\small
\caption{Results for the Multiscale MCMC method at point A: $\overline{X} = 7.99\times 10^{-2}$, $R(0)=0.148$.}\label{tablemultiA}
	\begin{tabular}{l|r|r|r|r|r|r}
	\hline
	Method          & avg IACT & min IACT & max IACT & Total ESS     & MCSE ($\times 10^{-3}$) & 95\%-C.I. ($\times 10^{-2}$)\\
	\hline
	AR(p) Fitting   &  608.75  &  583.80  &  659.18  & 13798.67 &  $3.27$ & $[7.35, 8.64]$ \\
   Batch means     & 2251.33  & 2121.73  & 2392.01  &  3731.14 &  $6.29$ & $[6.76, 9.22]$ \\
	Bartlett window & 2474.60  & 2326.22  & 2642.71  &  3394.49 &  $6.60$ & $[6.70, 9.29]$\\
	Tukey window    & 2553.68  & 2400.97  & 2723.57  &  3289.38 &  $6.70$ & $[6.68, 9.31]$\\
	ESS-Bulk 2      & 6302.90  & 4187.69  & 9191.10  &  1332.72 & $10.35$ & $[5.93, 10.06]$\\
	\hline
	ESS-Bulk 1      & \multicolumn{3}{|r|}{7818.50}  &  1074.38 & $11.72$ & $[5.70, 10.29]$\\
	ESS-Bulk 3      & \multicolumn{3}{|r|}{6904.01}  &  1216.68 & $11.02$ & $[5.83, 10.15]$\\
	\hline
	\end{tabular}
	\end{table*}

	\begin{table*}[t]
	\small
	\caption{Results for Two-stage Delayed-Acceptance method at point A: $\overline{X}=-0.177$, $R(0)=0.132$}\label{tabletwoA}
	\begin{tabular}{l|r|r|r|r|r|r}
	\hline
	Method          & avg IACT  & min IACT &  max IACT & ESS     & MCSE ($\times 10^{-3}$) & 95\%-C.I.\\
	\hline
	AR(p) Fitting   &   4486.35 &  3810.12 &   5313.07 & 4279.65 &  $5.55$ & $[-0.188, -0.166]$\\
	Batch means     &  13634.94 & 11052.17 &  15228.81 & 1408.15 &  $9.67$ & $[-0.196, -0.158]$ \\
	Bartlett window &  15242.15 & 12504.67 &  16874.96 & 1259.67 & $10.23$ & $[-0.197, -0.157]$\\
	Tukey window    &  15468.85 & 12636.01 &  17158.25 & 1241.20 & $10.30$ & $[-0.197, -0.157]$\\
	ESS-Bulk 2      & 133025.25 & 72252.01 & 202493.80 &  144.33 & $30.21$ & $[-0.236, -0.118]$\\
	\hline
	ESS-Bulk 1      &  \multicolumn{3}{|r|}{852599.36} &   22.52 & $76.48$ & $[-0.327, -0.027]$\\
	ESS-Bulk 3      &  \multicolumn{3}{|r|}{811319.66} &   23.67 & $74.60$ & $[-0.323, -0.031]$\\
	\hline
	\end{tabular}
	\end{table*}

	\begin{table*}[t]
	\small
\caption{Results for the Multiscale method at point B: $\overline{X}= -0.904$, $R(0)=0.172$}\label{tablemultiB}
	\begin{tabular}{l|r|r|r|r|r|r}
	\hline
	Method          & avg IACT & min IACT & max IACT & ESS     & MCSE ($\times 10^{-3}$) & 95\%-C.I.\\
	\hline
	AR(p) Fitting   &  1880.64 &  1762.44 &  1991.41 & 4466.57 & $6.21$ & $[-0.916, -0.892]$ \\
	Batch means     &  5236.21 &  4551.93 &  6118.74 & 1604.22 & $10.36$ & $[-0.924,-0.883]$\\
	Bartlett window &  5685.55 &  4954.07 &  6397.66 & 1477.43 & $10.80$ & $[-0.925, -0.883]$\\
	Tukey window    &  5911.27 &  5156.25 &  6618.60 & 1421.01 & $11.01$ & $[-0.925, -0.882]$\\
	ESS-Bulk 2      & 10172.39 &  8005.61 & 14131.63 &  825.76 & $14.44$ & $[-0.932, -0.875]$\\
	\hline
	ESS-Bulk 1      & \multicolumn{3}{|r|}{10818.35} &  776.46 & $14.90$ & $[-0.933, -0.875]$\\
	ESS-Bulk 3      & \multicolumn{3}{|r|}{11225.81} &  748.28 & 15.17 & $[-0.933, -0.874]$\\
	\hline
	\end{tabular}
	\end{table*}	
		
	\begin{table*}[t]
	\small
	\caption{Results for the Two-stage Delayed-Acceptance method at point B: $\overline{X} = -0.695$, $R(0)=0.152$}\label{tabletwoB}
		\begin{tabular}{l|r|r|r|r|r|r}
	\hline
	Method          & avg IACT   & min IACT &  max IACT & ESS    & MCSE ($\times 10^{-3}$)& 95\%-C.I.\\
	\hline
	AR(p) Fitting   &   17394.88 & 13447.21 &  21938.85 & 1103.77 & $11.75$ & $[-0.719, -0.672]$\\
	Batch means     &   21387.26 & 15800.95 &  29755.05 &  897.73 & $13.03$ & $[-0.721,-0.670]$ \\
	Bartlett window &   24113.74 & 17888.05 &  33481.97 &  796.23 & $13.84$ & $[-0.723, -0.668]$\\
	Tukey window    &   24422.48 & 18099.03 &  33835.66 &  786.16 & $13.92$ & $[-0.723, -0.668]$\\
	ESS-Bulk 2      &  188807.16 & 77853.86 & 291505.31 &  101.69 & $38.72$ & $[-0.771, -0.620]$\\
	\hline
	ESS-Bulk 1      &  \multicolumn{3}{|r|}{1369850.73} &   14.02 & 104.29 & $[-0.900, -0.491]$ \\
	ESS-Bulk 3      &   \multicolumn{3}{|r|}{883483.05} &   21.73 &  83.76  & $[-0.860, -0.531]$\\
	\hline
	\end{tabular}
	\end{table*}

Those MCMC methods use different prior distributions: The first method uses a Gaussian Random Field, and the second uses a patch of local Gaussian Random Fields. This means that the posterior distributions may not be the same. However, the study of their IACT or ESS may still give us an idea of how efficiently they are reaching those distributions. 

Each method was run with 4 independent chains in parallel, and all reported data is on $\eta$, the log permeability field. The stopping criterion for the methods was based on the marginal PSRF curves and their ESS. Our intention was to follow the suggestion in \cite{vehtari2021rank} and wait for the chains to reach an ESS of approximately 400 in all estimators and each chain. For the Multiscale method, we stopped at 2.6M iterations, and used the initial 500k iterations as burn-in. The Delayed Acceptance method took much longer to converge. Even after 5.8M iterations, the PSRF at some points remained stubbornly around 1.2, and some ESS estimators were also too slow to go up. We then decided to stop the simulation there, well short of the threshold of 400. As burn-in, the initial 1M iterations were used. 

Of the 256 points in the discretization of $\Omega = [0,1]\times [0,1]$, two were chosen for detailed analysis. The first, $A = (0.03125, 0.03125)$, is a point that has good convergence properties. The second, $B = (0.65625, 0.90625)$, is the point in which the Delayed-Acceptance method had the worst PSRF among the entire grid. 

\begin{figure*}[t]
	\centering
	\includegraphics[scale=\scaleval]{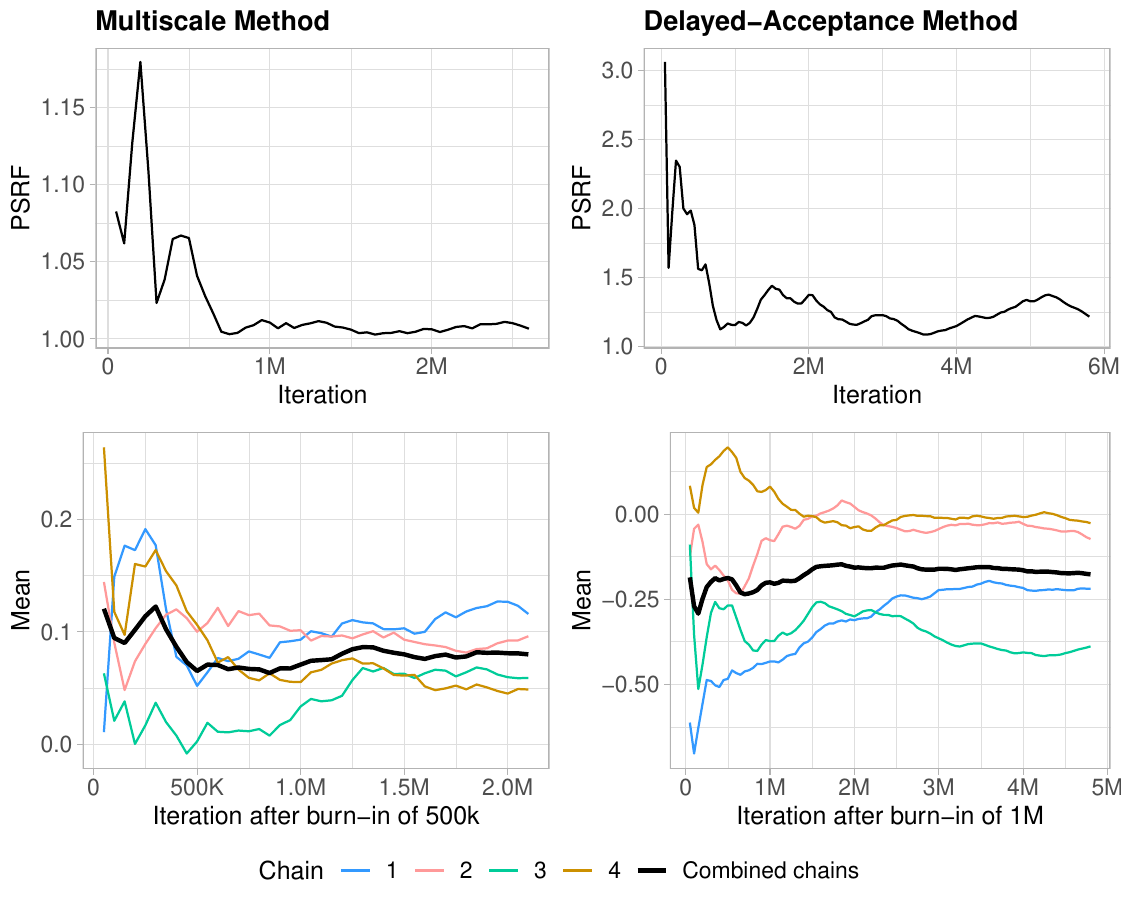}
	
	\vspace{0.3cm}
   \includegraphics[scale=\scaleval]{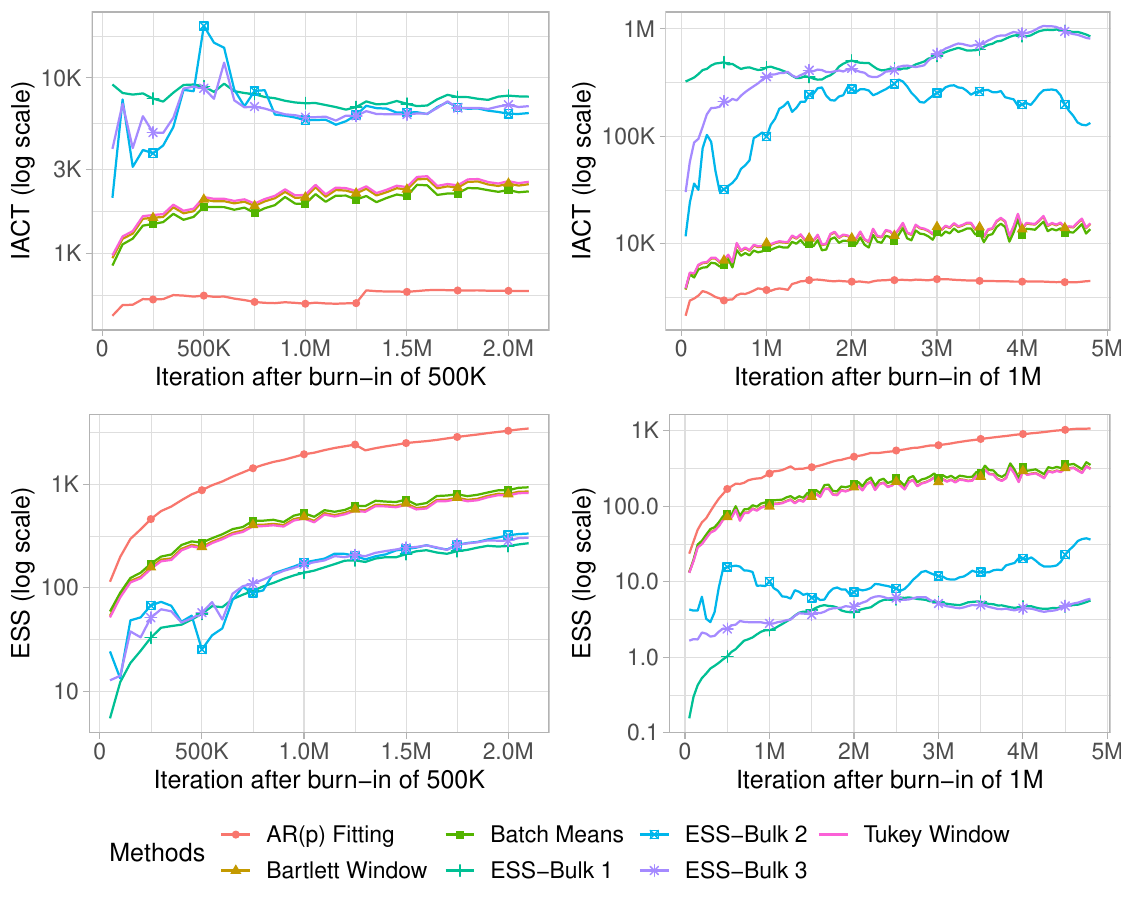}

	\caption{Results for point A. On the left we have the plots for the Multiscale Method, and those for the Delayed-Acceptance method are on the right. For the IACT and ESS plots, estimates for individual chains were averaged before plotting. Different scales were used to preserve the details of the plots}\label{ellipticfigsA}
	\end{figure*}

\begin{figure*}[t]
	\centering
	\includegraphics[scale=\scaleval]{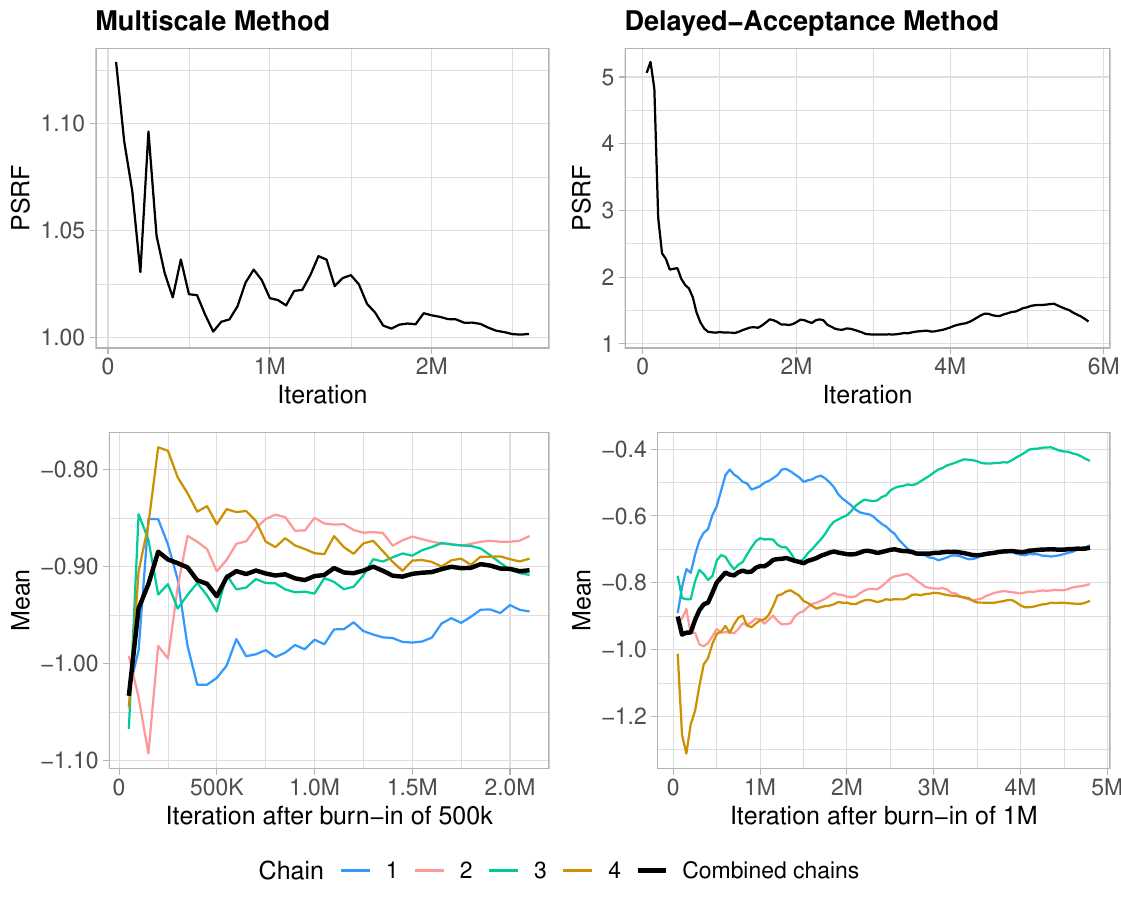}
	
   \vspace{0.3cm}
   \includegraphics[scale=\scaleval]{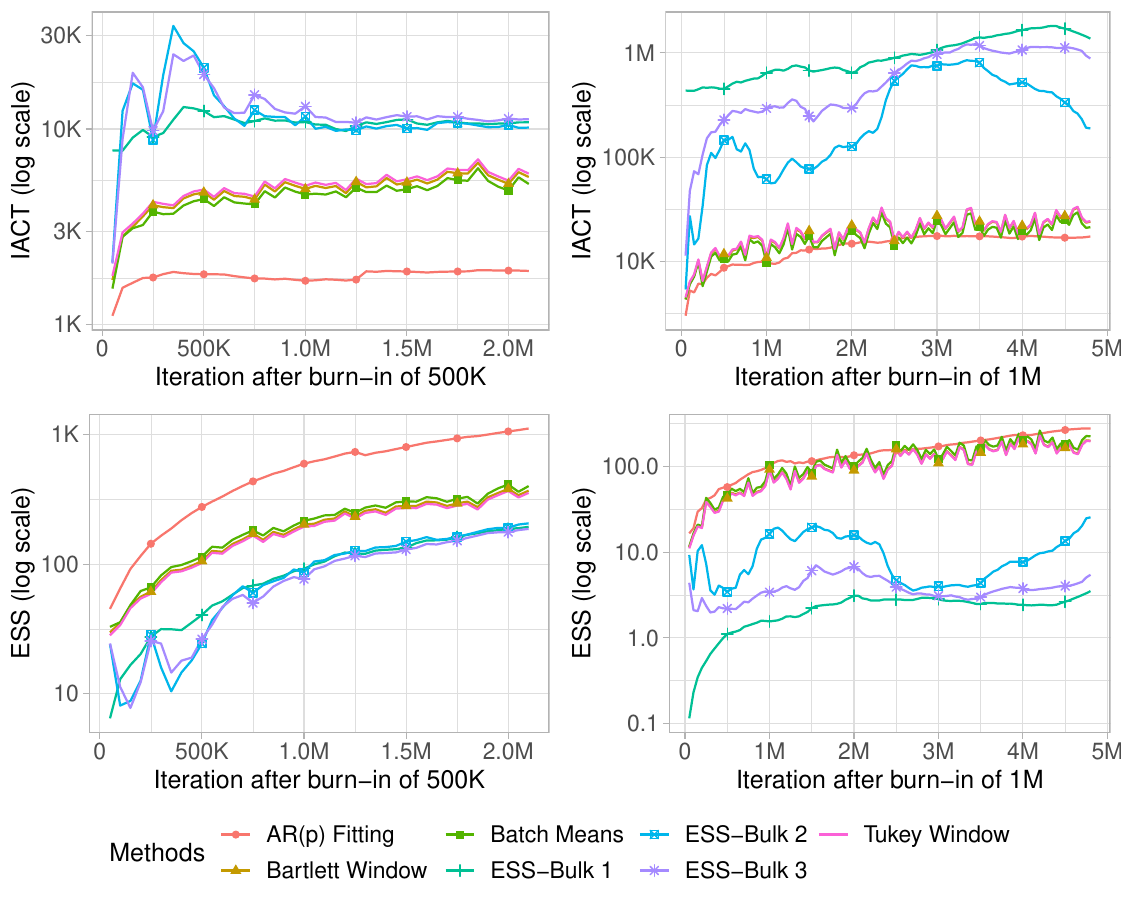}

	\caption{Results for point B. On the left we have the plots for the Multiscale Method, and those for the Delayed-Acceptance method are on the right. For the IACT and ESS plots, estimates for individual chains were averaged before plotting. Different scales were used to preserve the details of the plots}\label{ellipticfigsB}
	\end{figure*}

In \textit{Figures} \ref{ellipticfigsA} and \ref{ellipticfigsB}, we show the PSRF curves for the MCMC methods at points $A$ and $B$, respectively. Those were calculated with the \textit{coda} package. We note that we could have a different result using \textit{rstan}, since it rank-normalizes the sequence before estimating the PSRF (see comments in \textit{Section \ref{essb}}). Below those plots, we have the mean plots: They show the individual mean estimation for each chain, and then in bolder black the mean with all the chains put together. Then, we have the plots for the IACT and ESS estimations, using the same methods as in \textit{Section \ref{arexperiments}}. Apart from \textit{ESS-Bulk 1} and \textit{3}, the methods estimate the IACT for one chain at a time, so we averaged those results among the 4 chains.  

The final snapshot of the analysis using IACT and ESS is reported in \textit{Tables 3} to \textit{6}. The first column reports the IACT evaluated with the whole chain post-burnin. This amounts to 2,100,000 values for each chain in the Multiscale Method, and 4,800,000 for the Delayed Acceptance method. They were averaged for the methods that assess chains separately. The ESS column is the final value, using such IACT, on all 4 chains together. Because we did not remove any sample values as burn-in for the experiments we called \textit{ESS-Bulk 1}, the results for the IACT in those tables are for iterations 2,600,000 and 5,800,000, respectively. However, the ESS is calculated by dividing 2,100,000 and 4,800,000 by the IACT, so that they are comparable with the other estimators.

Finally, we report the MCSE and Confidence Interval for final value of the mean, putting all 4 chains together. We recall its formula, given in \textit{Section \ref{MCMCtheo}},
\begin{align*}
\text{MCSE} & = \frac{\widetilde{\gamma}}{\sqrt{N}}=\frac{1}{\sqrt{N}}\sqrt{\left(\sum_{k=-\infty}^{\infty}\rho(k)\right)R(0)}\\
&\approx\sqrt{\frac{\IACT\times R(0)}{N}}
\end{align*}
So, this will give us a total of 8,400,000 simulated values in the case of the Multiscale Method, and 19,200,000 values for the Delayed-Acceptance method.

\begin{figure}[h!]
\centering
\includegraphics[scale=\scaleval]{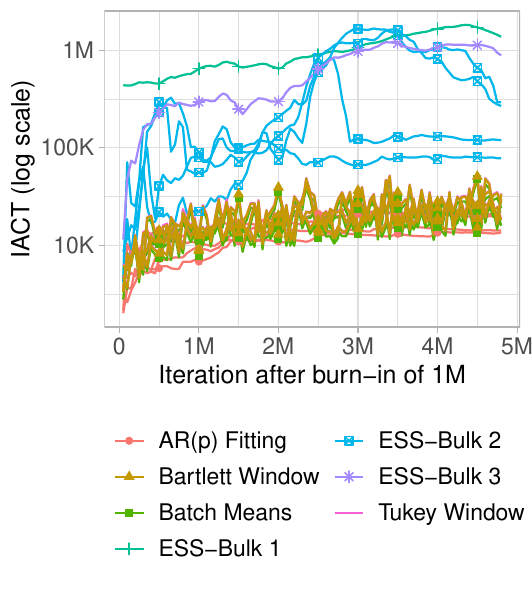}
\caption{IACT estimates for the Delayed-Acceptance method at point B. Each line is an estimate for a single chain. Since some methods combine the chains in their formula, they only have one line displayed}\label{iactallfig}
\end{figure}

How do we interpret all this data? Firstly, we recall from \textit{Section} \ref{arexperiments} that a single value of the ESS or IACT is not very useful, since its estimates vary quite a lot, and the estimators require an initial amount of sample values for them to be more accurate. Plotting both IACT and ESS is more useful, since they indicate whether those estimates have stabilized. For example, \textit{Figure \ref{iactallfig}} below is a complete picture of what is happening, by plotting the estimations for the individual chains. A good estimator should not deviate much between chains.

Perhaps more importantly, the estimators do not agree as to what the values of the IACT or ESS are. We conjecture that AR(p) Fitting is not an appropriate method, since for $p>1$ it is no longer a Markov Chain. Moreover, by the discussion in \textit{Section} \ref{essb}, we conjecture that the ESS-Bulk method significantly overestimates the IACT and, in turn, underestimates the ESS, for the reasons pointed out there: The PSRF doubly penalizes the estimator, and rank-normalization does the same. And although removing the burn-in values improved the estimations on \textit{ESS-Bulk 3} in comparison to \textit{ESS-Bulk 1}, \textit{Figure \ref{iactallfig}} tells us an important detail: The variation between the estimates on each chain is just too large for any conclusion to be made.    

\section{Conclusion}


From the experiments with AR(1) models in \textit{Section \ref{arprocess}}, we saw that the variance of the IACT estimate decreased very slowly with $N$, if at all, and thus less than the squared decay required to make the ESS estimations, at least in practice, consistent (see the beginning of \textit{Section \ref{estimateESS}}). That implies that the variance of the ESS estimate increases with $N$. Informally, the estimates for the ESS will get worse as $N$ increases (see \textit{Figures \ref{plotsar5k}} and \textit{\ref{plotsar50k}}). So, unless better estimators for the IACT are devised, we do not recommend using the ESS to decide when to stop the MCMC simulation of a complicated model. 

Also, the standard deviation of the IACT estimates can be of the same order of magnitude as its value. The consequence of this is that if two different MCMC methods have the IACTs of, say, 3000 and 5000, then they are indistinguishable. Moreover, reporting or observing just one final value of the IACT or ESS does not tell the full story: The IACT must itself ``converge'', and plotting it may help show that it is still growing or fluctuating. Related to this is the observation that the use of a small chain will likely underestimate the IACT. 


While the experiments with \textit{AR(1)} chains showed that the methods to calculate the IACT agreed in the order of magnitude and would all eventually converge to the true value, the same did not occur in the experiments with the inverse problem. We saw distinct orders of magnitude for the ESS depending on the estimator. For this type of problems, the methods using the ESS-Bulk and PSRF are suspected to severely underestimate the true ESS, and give an overly pessimistic picture of how the MCMC is converging. This will prompt its user to run the chains for far too many extra iterations than necessary -- and as stressed in that Section, each iteration is computationally expensive.


It doesn't even make sense to use the ESS as conservative estimate: Its lower bound can be as bad as zero. This is what we found in the simulation of the inverse problem, where the ESS of a single chain was estimated to be 5 even after 5 million iterations. Moreover, requiring the ESS to be ``large" before it can be trusted does not make sense either, because it is not, at least in practice, consistent.

One can also be tempted to also use the ESS to justify when an MCSE estimate is good enough, but there are better options. For example, the simplest estimator we discussed, the Batch Means method, has a more mature theoretical background that doesn't require the ESS in its arguments \cite{flegal2010batch}.

As far as just the IACT is concerned, we suggest relying on traditional estimatators instead of the newer ESS-Bulk. And even so, we recommend plotting it and understanding that this number is of limited interpretability in complicated models. It should therefore be used with care when comparing one MCMC method over another.

\vspace{1cm}
\noindent\textbf{Acknowledgements}

This material is based upon work supported by the National Science Foundation under Grant No. 2401945. Any opinions, findings, and conclusions or recommendations expressed in this material are those of the authors and do not necessarily reflect the views of the National Science Foundation.

The work of F.P. is also partially supported by The University of Texas at Dallas Office of Research and Innovation through the SPARK program. The author L.S. thanks the School of Natural Sciences and Mathematics and Department of Mathematical Sciences at the University of Texas at Dallas for the travel grant to present this work at the SIAM UQ24 Conference.

Both authors thank Dr. Arunasalam Rahunanthan and the Ohio Central State University for access to their GPU cluster. 

\bibliographystyle{elsarticle-num} 
\bibliography{mainbib.bib}

\appendix
\section{Remaining plots for the empirical distributions of the estimators for the ESS and IACT using AR(1) sequences}

\begin{figure*}[t]
\includegraphics[scale=\scaleval]{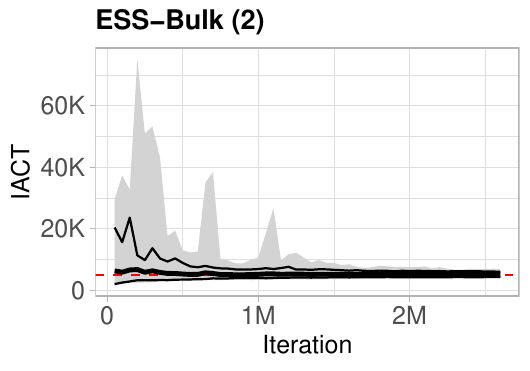}
\includegraphics[scale=\scaleval]{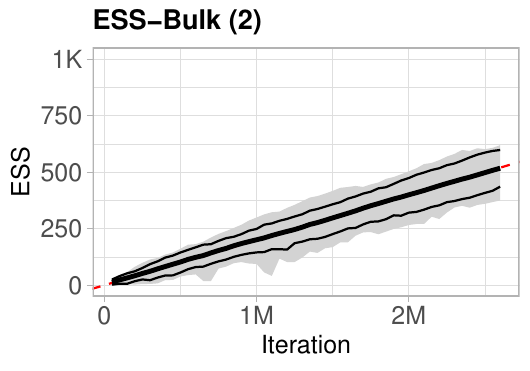}

\includegraphics[scale=\scaleval]{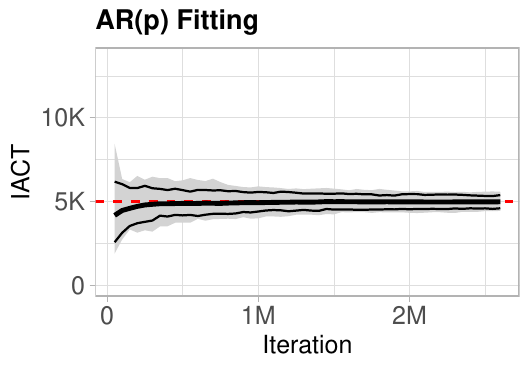}
\includegraphics[scale=\scaleval]{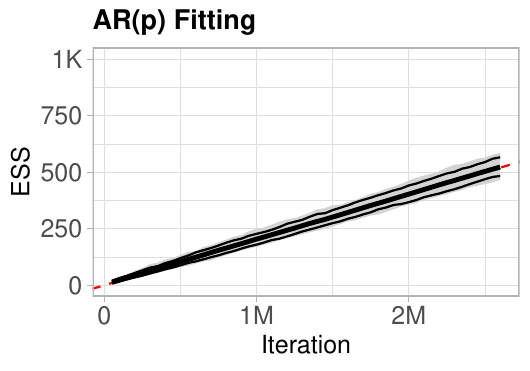}

\includegraphics[scale=\scaleval]{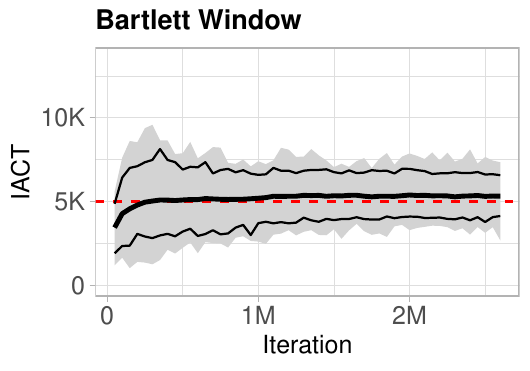}
\includegraphics[scale=\scaleval]{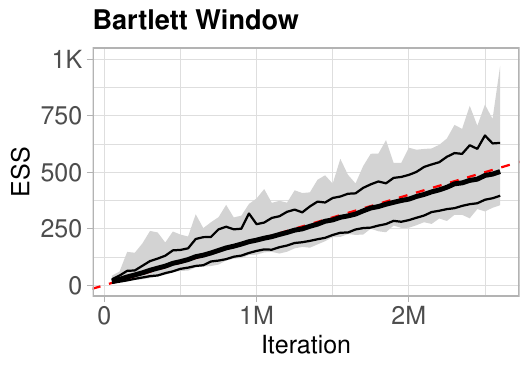}
 
\includegraphics[scale=\scaleval]{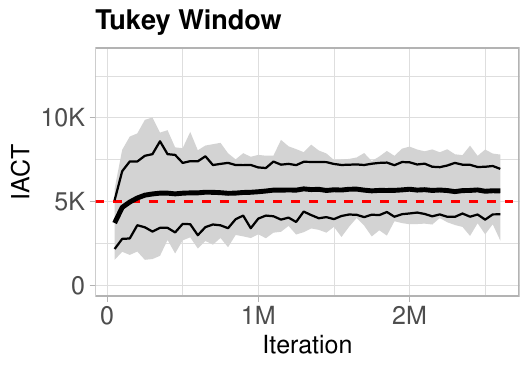}
\includegraphics[scale=\scaleval]{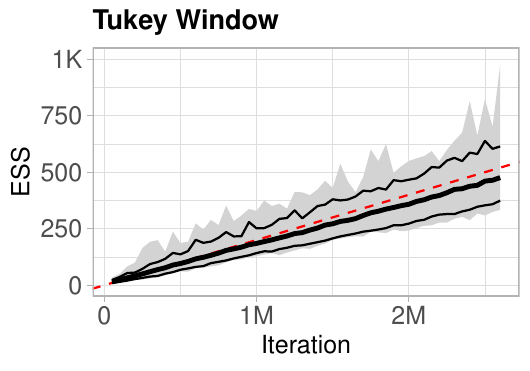}
\caption{IACT$=$5,000}\label{}
\end{figure*}

\begin{figure*}[t]
\includegraphics[scale=\scaleval]{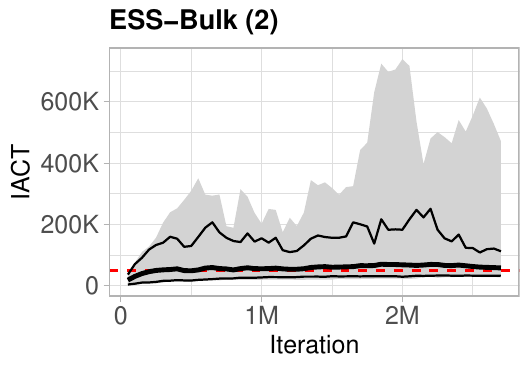}
\includegraphics[scale=\scaleval]{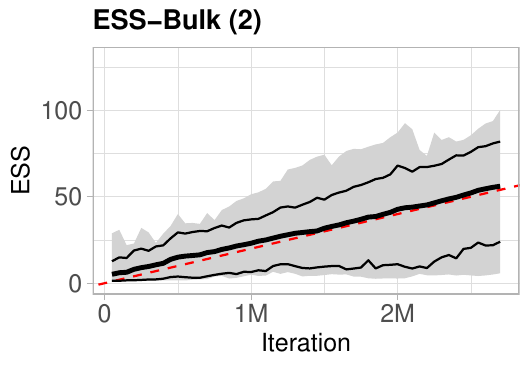}

\includegraphics[scale=\scaleval]{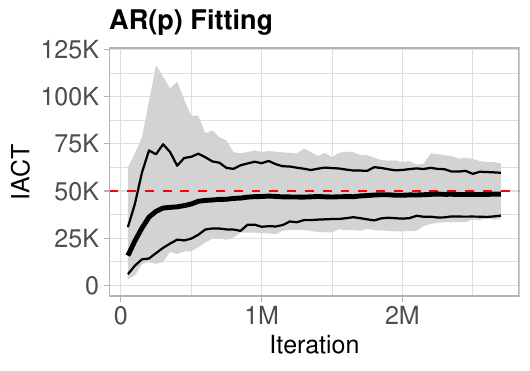}
\includegraphics[scale=\scaleval]{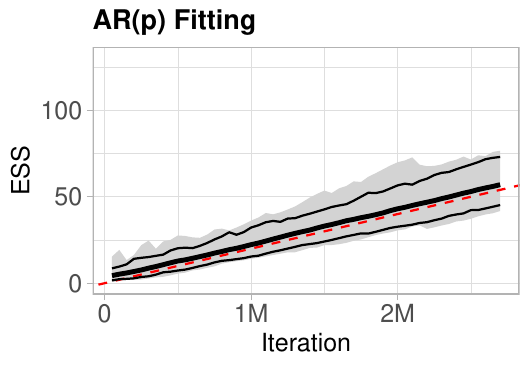}
 
\includegraphics[scale=\scaleval]{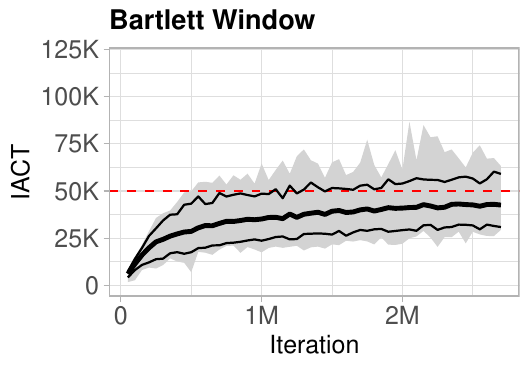}
\includegraphics[scale=\scaleval]{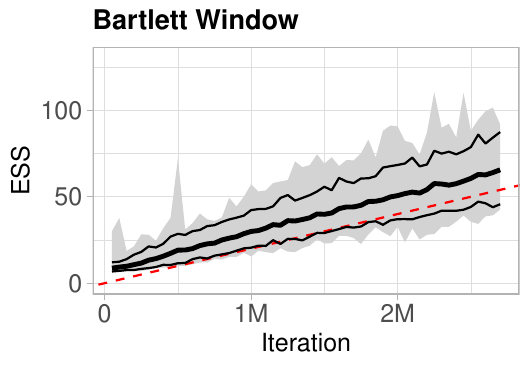}

\includegraphics[scale=\scaleval]{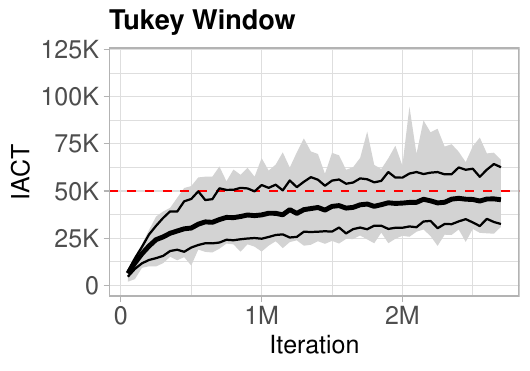}
\includegraphics[scale=\scaleval]{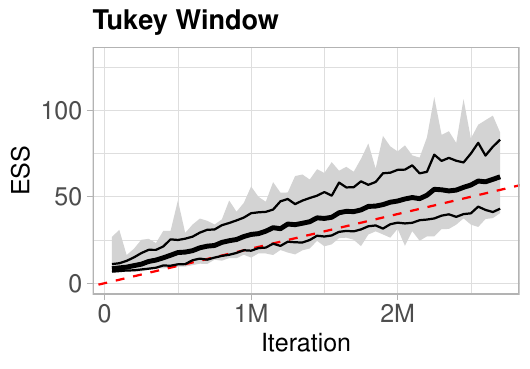}
\caption{IACT$=$50,000}\label{}
\end{figure*}

\end{document}